\documentclass{article}

\usepackage{icrctc07}

\title{Observations of the Pulsar PSR B1951+32 with the Solar Tower Atmospheric Cherenkov Effect Experiment}
\shorttitle{STACEE Observations of PSR B1951+32}
\authors{J. Kildea$^{1,2}$,
J. Zweerink$^3$, J. Ball$^3$, J.E. Carson$^{3,4}$, C. E. Covault$^5$, D.D. Driscoll$^5$,
P. Fortin$^6$, D. M. Gingrich$^{7,8}$,
D. S. Hanna$^1$, A. Jarvis$^3$, T. Lindner$^{1,9}$, 
C. Mueller$^1$, R. Mukherjee$^6$,
R. A. Ong$^3$, K. Ragan$^1$, D. A. Williams$^{10}$}
\shortauthors{Ball et al.}
\afiliations{
		$^{1}$  Department of Physics, McGill University, 
		     Montreal, QC H3A 2T8,\\
		$^{2}$  now at Fred Lawrence Whipple Observatory,
		     Amado, AZ 85645,\\
		$^{3}$  Department of Physics and Astronomy, University of California,
		     Los Angeles, Los Angeles, CA 90095,\\
		$^{4}$  now at Stanford Linear Accelerator Center, 
		     Menlo Park, CA 94025,\\
		$^{5}$  Department of Physics, Case Western Reserve University, 
           Cleveland, OH 44106,\\
		$^{6}$  Department of Physics and Astronomy, Barnard College, 
		     Columbia University, New York, NY 10027,\\
		$^{7}$  Department of Physics, University of Alberta,
		     Edmonton, AB T6G 2G7,\\
		$^{8}$  also at TRIUMF, Vancouver, BC V6T 2A3 Canada,\\
		$^{9}$  now at Department of Physics and Astronomy,
		University of British Columbia,
		     Vancouver, BC V6T 1Z1\\
		$^{10}$  Santa Cruz Institute for Particle Physics, 
		     University of California, Santa Cruz, Santa Cruz, CA 95064,\\
}
\email{jkildea@cfa.harvard.edu}

\abstract{
We present the analysis and results of 12.5 hours of high-energy
gamma-ray observations of the EGRET-detected pulsar PSR B1951+32 using
the Solar Tower Atmospheric Cherenkov Effect Experiment (STACEE).
STACEE is an atmospheric Cherenkov detector, in Albuquerque, New
Mexico, that detects cosmic gamma rays using the shower-front-sampling
technique. STACEE's sensitivity to astrophysical sources at energies
around 100 GeV allows it to investigate emission from gamma-ray
pulsars with expected pulsed emission cutoffs below 100 GeV. We
discuss the observations and analysis of STACEE's PSR 1951+32 data,
accumulated during the 2005 and 2006 observing seasons.
}

\begin{document}
\maketitle

\section{Introduction}
Young energetic pulsars (rapidly rotating, highly magnetized neutron stars, that produce non-thermal photons) remain as yet the only Galactic objects unambiguously detected within the EGRET energy range. To date, more than 1500 pulsars have been observed at radio energies \cite{manchester05}. About 70 of these are seen in X-rays \cite{kaspi06} but only 6 (Crab, Geminga, Vela, PSR B1951+32, PSR1706-44, and PSR B1055-52) were detected by EGRET \cite{nolan96}. Of these, only one, PSR B1951+32, has been observed up to $\sim$20~GeV without any apparent cutoff in its differential energy spectrum \cite{ramanamurthy95}. Accordingly, PSR 1951+32 is often considered the best pulsar candidate for pulsed-emission detection in the very-high-energy (VHE) regime; i.e.\ at energies above $\sim$50~GeV.

As is the case for most pulsars, PSR B1951+32 was first detected at radio energies. It was discovered with a 39.5~ms period in the radio synchrotron nebula CTB 80 \cite{kulkarni98}. From the radio observations it was deduced that the pulsar has a characteristic age of 1.1$\times$10$^5$~yr with a surface magnetic field of 4.9$\times$10$^{11}$~G, and a rotational energy loss rate of 3.7$\times$10$^{36}$~ergs~s$^{-1}$. It has also been observed in X-rays. While the pulsar emits a single pulse at radio frequencies and in X-rays, the EGRET observations display a double-pulsed profile with neither of the gamma-ray peaks coinciding with the radio peak. Interestingly, there is no evidence for any interpeak emission in the gamma-ray data \cite{ramanamurthy95}. 

At TeV energies, only upper limits on the pulsed emission, and on the emission from the surrounding synchrotron nebula, exist \cite{srinivasan97}. A recent report by the MAGIC collaboration \cite{albert07} presents observations above 75~GeV with no evidence for pulsed emission. 

Models that attempt to explain the non-thermal high-energy pulsed emission from pulsars generally fall into one of two broad categories; the Polar Cap \cite{muslimov03} or Outer Gap \cite{hirotani01} models. While both models can explain the observed gamma-ray emission at EGRET energies, they differ in their predictions for detectable emission above 20~GeV. In both scenarios the pulsed emission is attributed to particle acceleration in the pulsar's magnetosphere, with spectral cutoffs predicted in the 20-100~GeV energy range. 

The Polar Cap model localizes the emission site to a region close to the magnetic poles of the neutron star, where the magnetic field is strong. The Outer Gap model, on the other hand, contends that gamma-ray production occurs far from the neutron star surface in a region of relatively weak magnetic field, in so called ``outer gaps'' near the null surface of the outer magnetosphere. 

The emission sites dictate the energy of the expected spectral cutoff, insofar as the maximum energy of the curvature-radiated photons escaping the magnetosphere is limited by pair-production in the pulsar's magnetic field. Since the magnetic field is stronger near the polar cap, the Polar Cap model anticipates a lower-energy cutoff than the Outer Gap model. 

Any detection of TeV emission would clearly favor the Outer Gap model and would thereby significantly contribute to our understanding of pulsar emission processes. Thus far, no pulsed TeV emission has been detected from any pulsar, although ever-lower upper limits are constraining the emission models. As a gamma-ray observatory operating at the lower end of the VHE regime (around 100 GeV) the Solar Tower Atmospheric Cherenkov Effect Experiment (STACEE) is suited to pulsar observations. STACEE observations of PSR B1951+32, undertaken in 2005 and 2006, are reported here. 

\section{STACEE}
STACEE is a shower-front-sampling atmospheric Cherenkov telescope that uses the facilities of the National Solar Thermal Test Facility (NSTTF) in Albuquerque, New Mexico \cite{Gingrichetal2004}. The NSTTF is a solar energy research facility incorporating a central receiver tower and an array of heliostats (solar mirrors). STACEE uses secondary mirrors, in the central receiver tower, to focus Cherenkov light from air showers, that is reflected by the heliostats, onto cameras having a total of 64 photomultiplier tubes (PMTs). A one-to-one mapping between heliostats and PMTs allows the Cherenkov shower-front to be sampled independently at 64 different locations, spread over an area of $\sim$2$\times10^4$~m$^2$, on the heliostat field. The large reflecting surface provided by the 64 heliostats, each with an area of 37~m$^2$, allows STACEE to operate with an energy threshold around 100~GeV. 

STACEE uses a custom-built 2-level trigger system \cite{Gingrichetal2004} to select Cherenkov events from amongst the background of night-sky light fluctuations. In the event of a Cherenkov trigger, amplified and AC-coupled signals from the PMTs are recorded, together with a GPS timestamp (1$\mu$s resolution), using 8-bit Flash Analog to Digital Convertors (FADCs), one per PMT. The FADCs provide important temporal and intensity information, at a sampling rate of 1~GS/s, which is fully utilized in the offline data analysis procedure. 
 

\section{Observations of PSR B1951+32}
STACEE observations of PSR B1951+32 were undertaken during clear moonless nights in the periods June-July 2005, September 2005 and June 2006. The declination of PSR B1951+32 is favorable for high-elevation observations by STACEE and most data were recorded at an elevation of 82.5 degrees. 

Observation runs of 28 minute duration were conducted in an on-source-only mode. This is different from STACEE's normal ``ON/OFF'' observing mode in which observations of both the source region (ON) and a control region of sky (OFF) are recorded. For PSR B1951+32, on-source-only observations were considered appropriate, given the lack of interpeak gamma-ray emission observed by EGRET \cite{ramanamurthy95} and the desire to maximize the amount of data recorded. Should the source emit gamma rays above 100~GeV within the sensitivity of STACEE, they would be detectable as an excess number of events within the EGRET pulsed region (phases ranges 0.12-0.22 and 0.48-0.74), compared to the number of events outside the pulsed region, assuming that the EGRET pulse profile is maintained at TeV energies. 

A total of 15.1 hours of PSR B1951+32 data were recorded by STACEE during 2005 and 2006.

\section{Data Quality Selection}
Before STACEE data are analyzed for the presence of a gamma-ray signal, data quality selection is performed to remove sections of data flagged as unusable due to hardware malfunctions, unstable atmospheric conditions or transient terrestrial light contamination. Heliostat and FADC malfunctions are logged each night by the heliostat and FADC control software respectively, and later merged with the Cherenkov data offline. Hence, data contaminated by hardware malfunctions are easily identified and removed. 

Data contaminated by unstable atmospheric conditions or transient background light are usually identified by cross-checking the level-1 trigger rates (the rates at which clusters of 8 heliostats register coincident threshold-crossing light pulses) between ON and OFF data. Under stable conditions, the ON and OFF level-1 trigger rates should behave similarly. For the PSR~B1951+32 observations reported here, a modified data quality checking procedure was necessary since OFF data were not available for comparison. The following procedure was adopted. 

For each data run recorded, the level-1 trigger rates were binned in 10-second time slices and lightcurves of the rates as a function of time produced. Second-order polynomial fits to the lightcurves were used to characterize the elevation-dependence of the trigger rates over the course of the data run. The residual difference between the true rate and the fitted rate in each time slice was then examined and slices with a residual more than 3 times the RMS residual were flagged as bad. Time slices for which three or more clusters were labeled as bad were cut from the data set, together with 25-second adjacent time sections. In this way, 1 minute periods of data with anomalous level-1 trigger rates---indicative of unstable weather or local light contamination---were excluded from the final data analysis. 


Data quality selection removed approximately 2.6 hours of PSR B1951+32 data, such that the final data set used in the gamma-ray and temporal analyses was 12.5 hours of clean data. 

\section{Data Analysis}
Data analysis for STACEE is a two-stage process. In stage 1, fixed and dynamic time corrections are applied to the pulse times registered for each PMT. Fixed delays are necessary to account for the measured photon time-of-flight delays from the individual heliostats to the PMTs and for the signal propagation delays between the PMTs and the central trigger system. Dynamic delays account for the different and changing arrival times of the Cherenkov shower-front at the heliostats. 


In stage 2, the time-corrected FADC data are used to determine the shower core---the point at which the shower's axis intersects the Earth's surface. A template-fitting approached is used. PMT-charge templates are compiled using showers simulated over a large range of zenith angles, azimuth angles, and core locations. By finding the template that best matches a particular event, an approximate core location for that event is obtained. Core resolution using this method is about 15~m, according to simulations. 

The estimated shower core location is used as the starting point of the ``grid ratio'' analysis---the analysis routine used by STACEE for gamma/hadron discrimination \cite{kildea05,lindner07}. The grid ratio essentially provides a measure of the sphericity of the shower-front. Since gamma-ray showers with energies in the STACEE operating range typically have spherical shower fronts, whereas hadron showers are irregular and non-spherical, a cut on the sphericity of the shower-front allows for a significant reduction of the large hadronic background.

\section{Temporal Analysis}
The arrival times of the candidate gamma-ray events, as registered by the STACEE GPS clock, were transformed to the solar system barycenter using the JPL DE200 Planetary and Lunar Ephemeris. Two independent barycentering algorithms were employed with excellent agreement. The pulsar phase of each event at the solar system barycenter, with respect to the PSR~B1951+32 radio ephemeris, was calculated and a phaseogram produced.

\begin{figure}
\begin{center}
\resizebox{17pc}{!}{\includegraphics{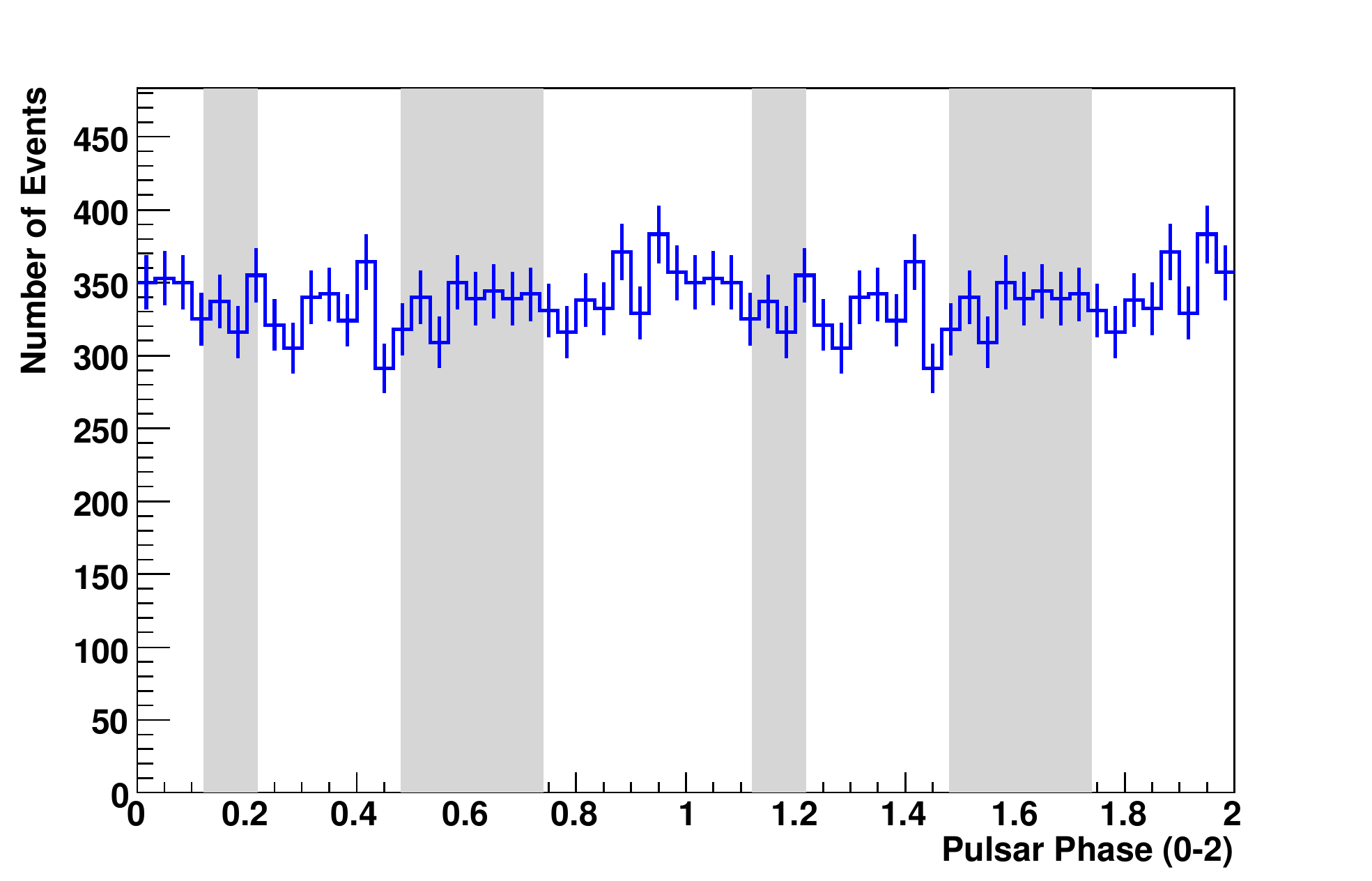}}
\caption{Phaseogram of candidate STACEE gamma-ray events for PSR B1951+32 folded using the
pulsar's radio ephemeris. The shaded regions show the positions of the main pulse (phase 0.12--0.22), and the interpulse (phase 0.48--0.74) as seen by EGRET \cite{ramanamurthy95}}

\label{fig:optical}
\end{center}
\end{figure}

To test the barycentering algorithms, an optical signal of the Crab pulsar, as seen using three PMTs set up at STACEE in a special arrangement, was analyzed. The resulting optical lightcurve, showed clear evidence of a pulsed signal phase-aligned with the Crab pulsar radio pulsations \cite{fortin05}. This demonstrated the validity of the barycentering software and also STACEE's timing electronics. 

\section{Results and Conclusions}
No evidence for pulsed gamma-ray emission above 117 GeV was found in the
selected gamma-ray data set used in this work (Figure 1). To calculate flux upper limits,
we used the method of Helene \cite{helene83} to estimate 99.0\% upper limits for excess
events within the pulsed phase profile seen by EGRET at lower energies \cite{ramanamurthy95}.
That is, emission is assumed to occur in the phase range of the main pulse,
phase 0.12--0.22, and the interpulse, phase 0.48--0.74. A differential flux upper
limit of 4.52~$\times 10^{-6}$~MeV~cm$^{-2}$~s$^{-1}$ was determined at the energy threshold of
117 GeV, by extrapolation of the EGRET spectrum to STACEE energies.

\bigskip
{\bf Acknowledgements:}
Many thanks go to the staff of the National Solar Tower Test Facility, who
have made this work possible.
This work was funded in part by the US National Science Foundation, the
Natural Sciences and Engineering Research Council of Canada, Fonds Quebecois
de la Recherche sur la Nature et les Technologies, the Research
Corporation, and the University of California at Los Angeles.

\end{document}